\documentclass[preprint,preprintnumbers]{revtex4}
\usepackage{amssymb}

\usepackage{amsmath}
\usepackage{graphicx}
\usepackage{dcolumn}
\usepackage{bm}

\input{tcilatex}

\begin{document}

\title{Voltage dependent conductance and shot noise in quantum microconstriction
with single defects. }
\author{Ye.S. Avotina$^{\left( 1\right) },$ A. Namiranian$^{\left( 2\right) }$, \
Yu.A. Kolesnichenko$^{\left( 1\right) }$ \\
$^{(1)}$B.Verkin Institute for Low Temperature Physics and Engineering,
National Academy of Sciences of Ukraine, 47 Lenin Ave., 310164 Kharkov,
Ukraine.\\
$^{(2)}$Physics Department, Iran University of Science and Technology,
Narmak, 16844, Teheran, Iran.}

\begin{abstract}
The influence of the interference of electron waves, which are scattered by
single impurities and by a barrier on nonlinear conductance and shot noise
of metallic microconstriction is studied theoretically. It is shown that the
these characteristics are nonmonotonic functions on the applied bias $V$.
\end{abstract}

\maketitle

\section{Introduction}

Single defects influence strongly to physical properties of mesoscopic
systems. Usually different defects arise during a manufacturing of
mesoscopic conductors and an investigation of its effect to the transport
characteristics has a practical significance. From other hand the study of
the effect of single defects to kinetic coefficients makes it possible to
obtain the most detailed information on the electron scattering processes
that is important for the basic science. Point contacts and quantum
microconstrictions (quantum wires) are one of the classes of mesoscopic
systems, which are wide investigated both theoretically and experimentally
(for review see \cite{Blanter,Agrait}). The electrical conductance $G$ of
such constrictions is proportional to the number $N$ of propagating electron
modes (the number of discrete energy levels $\varepsilon _{n}<$ $\varepsilon
_{F}$ of transverse quantization, $\varepsilon _{F}$ is the Fermi energy),
each one contributing an amount of $G_{0}=2e^{2}/h.$ The changing of the
contact diameter $d$ leads to the changing of the number of occupied levels $%
\varepsilon _{n}$ and the $G\left( d\right) $ suffers a $G_{0}$ stepwise
change. This effect is a manifestation of the quantum size effect in metals,
which first predicted by Lifshits and Kosevich in 1955 \cite{Lifshits}. The
scattering processes decrease a probability $T_{n}<1$ of the transmission of 
$n-th$ mode and the conductance at zero temperature $T=0$ and an applied
voltage $V\rightarrow 0$ is described by Landauer-Buttiker formula \cite
{Landauer,Buttiker}.

The shot noise is an important characteristic of the transport properties of
mesoscopic conductors \cite{Blanter,Agrait,Kogan}. It is originated from the
time-dependent current fluctuations. First Kulik and Omelyanchouk \cite
{KulOm} noticed that the shot noise in ballistic contacts vanishes in the
quasiclassical approximation, if there is no any electron scattering. In
quantum microconstriction these fluctuations arise from the quantum
mechanical probability of electrons to be transmitted through it. At $T=0$,
a bias at the contact $V\rightarrow 0$ and for low frequencies $\omega
\rightarrow 0$ the shot noise is given by \cite{Blanter} 
\begin{equation}
S\left( 0\right) =2eVG_{0}\sum_{n=1}^{N}T_{n}\left( 1-T_{n}\right) .
\label{S(0)}
\end{equation}
In perfect ballistic contacts, in which the transmission probability for
every mode $T_{n}=1,$ the shot noise is fully suppressed. However even in
adiabatic ballistic constrictions near the values of its diameter, at which
the highest energy levels $\varepsilon _{N}$ \ is close to $\varepsilon
_{F}, $ the probability $T_{N}<1$\cite{Bogachek}. According to Eq. (\ref
{S(0)}) at the small bias the shot noise is the linear function of the
voltage $V$.

The conductance of quantum microconstriction containing different types of
single defects has been investigated theoretically in the papers \cite
{Namir1}-\cite{Avotina}. The most remarkable features of manifestation of
electron scattering process in mesoscopic constrictions with only few
point-like defects are: (i) the effect of quantum interference between
directly transmitted through the contact an electron wave and electron waves
scattered by the defects and a barrier in the contact and (ii) the
dependence an electron scattering amplitude on a position of the defect
inside the constriction. First effect results in the nonmonotonic dependence
of the point-contact conductance on the applied bias, which was observed in
the experiments \cite{Untiedt,Ludoph} and theoretically considered in the
papers \cite{Namir1,Ludoph}. Recently new experimental observation of
conductance oscillations in quantum contact was reported in Ref.\cite{Kempen}%
. The second one is the reason for a size dependence of the Kondo anomaly of
the quantum conductance \cite{Namir2,Zarand}. This dependence is due to a
non-homogeneity of the local density of the electron states across of the
diameter of microconstriction. In the paper \cite{Maslov} based on a
numerical simulation the influence of ''dirty'' banks on the conductance of
quantum point contact was considered and authors had predicted the
suppression of conductance fluctuations near the edges of the steps of the
function $G\left( d\right) .$ This effect have been experimentally observed
in Ref.\cite{Ludoph} and explained by the decreasing of the interference
terms in the conductance, if the contact diameter $d$ is closed to the jump
of the $G\left( d\right) $.

The most important feature of the ballistic microconstriction is the
splitting of the Fermi surface by applied voltage \cite{KOSh}. Effectively,
there are two electronic beams moving in opposite directions with energies
differing at each point of the constriction by exactly the bias energy $eV.$
Because of this difference of electron energies $\varepsilon \pm \frac{eV}{2}%
,$ a value of a wave vector $k_{z}\left( \varepsilon \pm \frac{eV}{2}\right) 
$ along the constriction depends on $eV$ . The mentioned above the effect of
quantum interference between directly transmitted and scattered waves
defines by a relative phase shift $\Delta \varphi =2k_{z}\Delta z$ of wave
functions ($\Delta z$ is a distance between scatterers) and the dependence $%
k_{z}\left( \varepsilon \pm \frac{eV}{2}\right) $ results in oscillations of
transmission probabilities $T_{n}\left( V\right) $ as functions on $V$. In
this paper we consider an influence of such effect to the conductance and
shot \ noise in long quantum microconstrictions with few defects and the
potential barrier.

The paper is organized as follows. In Sec. II the model of microconstriction
and the basic equations are discussed. In Sec. III the voltage dependence of
conductance and shot noise is studied. The two cases are considered: single
impurity in the constriction with a barrier and two impurities in the
constriction without the barrier. The expressions for Green's function for
these cases are given. Also the results of numerical calculations are
presented in this section. We summaries our results in Sec. IV.

\section{Model of microconstriction and formulation of the problem}

\bigskip Let us consider the quantum microconstriction in the form of a long
channel with smooth boundaries and a diameter $2R$ comparable with the Fermi
wavelength $\lambda _{F}$ (Fig.1). A length of the channel $L$ is much
larger than $R.$ We assume that the channel is smoothly (over Fermi length
scale) connected with a bulk metal banks, to which the voltage $eV\ll
\varepsilon _{F}$ is applied. In a center of the constriction a potential
barrier $\left( U\left( z\right) =U\delta \left( z\right) \right) $ is
situated in vicinity of which there are few point-like defects in positions $%
\mathbf{r}_{i}$. The Hamiltonian of the system can be written as 
\begin{equation}
\widehat{H}=\frac{\widehat{\mathbf{p}}^{2}}{2m^{\ast }}+U\delta \left(
z\right) +g\sum_{i}\delta \left( \mathbf{r-r}_{i}\right) ,  \label{H}
\end{equation}
where $\widehat{\mathbf{p}}$ and $m^{\ast }$ are a momentum operator and an
effective mass of an electron, $g$ is a constant of electron-impurity
interaction ($g>0$, a repulsive impurity). In a ballistic channel without
the barrier and defects $\left( U=g=0\right) $ the wave functions and
energies of the eigenstates inside the channel can be separated to the
transversal and longitudinal parts with respect to the constriction axis $z$%
: 
\begin{eqnarray}
\Psi _{\alpha }\left( \mathbf{r}\right) &=&\frac{1}{\sqrt{L}}\psi _{\bot
\beta }\left( \mathbf{R}\right) e^{ik_{z}z};  \label{psi} \\
\varepsilon _{\alpha } &=&\varepsilon _{\beta }+\frac{\hbar ^{2}k_{z}^{2}}{%
2m^{\ast }},  \label{energ}
\end{eqnarray}
where $\alpha =\left( \beta ,k_{z}\right) $ is a full set of quantum numbers
consisting of two discrete quantum numbers $\beta =\left( m,n\right) ,$
which define the discrete energies $\varepsilon _{\beta }$\ of conducting
modes, and $k_{z}$ is the wave vector along the $z$ axis; $\mathbf{r=}\left( 
\mathbf{R,}z\right) .$ The transversal part $\psi _{\bot \beta }\left( 
\mathbf{R}\right) $ of \ the wave function satisfies to zero boundary
conditions at the surface of the constriction. The functions $\Psi _{\alpha
}\left( \mathbf{r}\right) $ are orthogonal and normalized.

According to definition the noise power spectrum is 
\begin{equation}
S_{ab}\left( \omega \right) =\frac{1}{2}\int dte^{i\omega t}\left\langle
\Delta \widehat{I}_{a}\left( t\right) \Delta \widehat{I}_{b}\left( 0\right)
+\Delta \widehat{I}_{b}\left( 0\right) \Delta \widehat{I}_{a}\left( t\right)
\right\rangle ,
\end{equation}
where $\Delta \widehat{I}_{a}\left( t\right) =\widehat{I}_{a}\left( t\right)
-I_{a};$ $\widehat{I}_{a}\left( t\right) $ is the current operator in the
right $(a,b=R)$ or left $\left( a,b=L\right) $ lead; $I_{a}=\left\langle 
\widehat{I}_{a}\right\rangle $ is the average current in the lead $a;$
brackets $\left\langle ...\right\rangle $ denotes the quantum statistical
average for a system at thermal equilibrium. In this paper we will only be
interested in zero frequency noise $S_{ab}\left( 0\right) .$ Note that due
to the current conservation $I\equiv I_{L}=$ $I_{R}$ we have $S\equiv
S_{LL}=S_{RR}=-S_{LR}=-S_{RL}.$

The general formula for a current $I$ through the quantum contact at a
arbitrary voltage was obtained by Bagwell and Orlando \cite{Bagwell} (see
also the book \cite{Datta}):

\begin{equation}
I=\frac{2e}{h}\int d\varepsilon T\left( \varepsilon ,V\right) \times \left(
f_{L}-f_{R}\right) ;
\end{equation}
where is the transmission coefficient of electrons through the constriction 
\begin{equation}
T\left( \varepsilon ,V\right) =Tr\left[ \widehat{t}^{\dagger }\left(
\varepsilon ,V\right) \widehat{t}\left( \varepsilon ,V\right) \right] ,
\label{T}
\end{equation}
and $\ f_{L,R}\left( \varepsilon \right) =f_{F}\left( \varepsilon \pm \frac{%
eV}{2}\right) $ is the distribution function of electrons moving in the
contact from the left $\left( f_{L}\right) $ or right $\left( f_{R}\right) $
bank; $f_{F}(\varepsilon )$ is the Fermi function, $\widehat{t}\left(
\varepsilon ,V\right) $ is a scattering matrix. In general case the function 
$T\left( \varepsilon ,V\right) $ depends on the applied voltage $V$ because
the electron scattering \ leads to the appearance of nonuniform electrical
field inside the constriction \cite{Lenstra}. This field must be found
self-consistently from the equation of electroneutrality. In an almost
ballistic microconstriction containing few scatterers and $\delta $-function
potential barrier of the small amplitude $U$ the mentioned electrical field
is small and we neglect its effect, assuming that the electrical potential
drops at the ends of the constriction.

In the same approximation the noise spectrum $S\left( 0\right) $ is given by 
\cite{Blanter,Agrait} 
\begin{eqnarray}
S\left( 0\right) &=&\frac{2e^{2}}{h}\int d\varepsilon \left\{ Tr\left[ 
\widehat{t}^{\dagger }\left( \varepsilon \right) \widehat{t}\left(
\varepsilon \right) \widehat{t}^{\dagger }\left( \varepsilon \right) 
\widehat{t}\left( \varepsilon \right) \right] \times \left[ f_{L}\left(
1-f_{L}\right) +f_{R}\left( 1-f_{R}\right) \right] \right.  \label{S} \\
&&\left. +Tr\left[ \widehat{t}^{\dagger }\left( \varepsilon \right) \widehat{%
t}\left( \varepsilon \right) \left( \widehat{I}-\widehat{t}^{\dagger }\left(
\varepsilon \right) \widehat{t}\left( \varepsilon \right) \right) \right]
\times \left[ f_{L}\left( 1-f_{R}\right) \right] +f_{R}\left( 1-f_{L}\right)
\right\} ;  \notag
\end{eqnarray}
where$\ \widehat{I}$ is the unit matrix. The first term in the Eq. (\ref{S})
corresponds to thermal fluctuations (the equilibrium, or Nyquist-Johnson
noise) and vanishes, if the temperature $T\rightarrow 0.$ The second part of
this equation remains finite at $T=0,$ if the bias is applied to the
constriction, and it describes the shot noise.

The calculation of the transport properties of the quantum constriction
comes to the determination of the scattering matrix $\widehat{t}\left(
\varepsilon \right) .$ Elements of scattering matrix $t_{\beta \beta
^{\prime }}$ can be expressed by means of the advanced Green's function $%
G^{+}\left( \mathbf{r,r}^{\prime };\varepsilon \right) $\ of the system \cite
{Fisher}: 
\begin{equation}
t_{\beta \beta ^{\prime }}\left( \varepsilon \right) =-\frac{i\hbar
^{2}k_{\beta ^{\prime }}}{m^{\ast }}G_{\beta \beta ^{\prime }}^{+}\left(
z,z^{\prime };\varepsilon \right) ;\;z\rightarrow -\infty ,z^{\prime
}\rightarrow +\infty ;  \label{t_beta}
\end{equation}
where 
\begin{equation}
k_{\beta }\left( \varepsilon \right) =\frac{1}{\hbar }\sqrt{2m^{\ast }\left(
\varepsilon -\varepsilon _{\beta }\right) }  \label{k_beta}
\end{equation}
is an absolute value of electron wave vector corresponding to the quantum
energy level $\varepsilon _{\beta };$ $G_{\beta \beta ^{\prime }}\left(
z,z^{\prime };\varepsilon \right) $ are components of the expansion of
Green's function on the full set of wave functions corresponding to the
transverse motion of electrons 
\begin{equation}
G^{+}\left( \mathbf{r,r}^{\prime };\varepsilon \right) =\sum_{\beta \beta
^{\prime }}\psi _{\bot \beta }\left( \mathbf{R}\right) \psi _{\bot \beta
^{\prime }}^{\ast }\left( \mathbf{R}^{\prime }\right) G_{\beta \beta
^{\prime }}^{+}\left( z,z^{\prime };\varepsilon \right) .  \label{G+}
\end{equation}
The matrix elements $t_{\beta \beta ^{\prime }}\left( \varepsilon \right) $
describes the transmission probabilities for carriers incident in channel $%
\beta $ in the left lead $L$ and transmitted into channel $\beta ^{\prime }$
in the right lead $R$. The Green's function satisfies the Dyson's equation: 
\begin{equation}
G\left( \mathbf{r,r}^{\prime };\varepsilon \right) =G_{b}\left( \mathbf{r,r}%
^{\prime };\varepsilon \right) +g\sum_{i}G_{b}\left( \mathbf{r,r}%
_{i};\varepsilon \right) G\left( \mathbf{r}_{i}\mathbf{,r}^{\prime
};\varepsilon \right) ,  \label{G_eq}
\end{equation}
where $G_{b}\left( \mathbf{r,r}^{\prime };\varepsilon \right) $ is the
Green' function of ballistic microconstriction with the barrier in the
absence of defects. It can be found from the equation 
\begin{equation}
G_{b}\left( \mathbf{r,r}^{\prime };\varepsilon \right) =G_{0}\left( \mathbf{%
r,r}^{\prime };\varepsilon \right) +U\int d\mathbf{R}^{^{\prime \prime
}}G_{0}\left( \mathbf{r;R}^{\prime \prime },z^{\prime \prime }=0;\varepsilon
\right) G_{b}\left( \mathbf{R}^{\prime \prime },z^{\prime \prime }=0;\mathbf{%
r}^{\prime };\varepsilon \right) ,  \label{Dyson_bar}
\end{equation}
where 
\begin{equation}
G_{0}^{+}\left( \mathbf{r,r}^{\prime };\varepsilon \right) =\sum_{\beta }%
\frac{m^{\ast }}{i\hbar ^{2}k_{\beta }}\psi _{\bot \beta }\left( \mathbf{R}%
\right) \psi _{\bot \beta }^{\ast }\left( \mathbf{R}^{\prime }\right)
e^{ik_{\beta }\left\vert z^{\prime }-z\right\vert }  \label{G0}
\end{equation}
is the Green's function in the absence of impurities and the barrier.
Substituting the expansions (\ref{G+}) and (\ref{G0}) into equation (\ref
{Dyson_bar}) and taking into account the orthogonality of functions $\psi
_{\bot \beta }\left( \mathbf{R}\right) $ for the coefficients $G_{b\beta
}^{+}\left( z,z^{\prime };\varepsilon \right) \delta _{\beta \beta ^{\prime
}}$ of $G_{b}^{+}\left( \mathbf{r,r}^{\prime };\varepsilon \right) $ in the
expansion (\ref{G+}) we obtain the algebraic equation 
\begin{equation}
G_{b\beta }^{+}\left( z,z^{\prime };\varepsilon \right) =\frac{m^{\ast }}{%
i\hbar ^{2}k_{\beta }}\left[ e^{ik_{\beta }\left\vert z^{\prime
}-z\right\vert }+Ue^{ik_{\beta }\left\vert z\right\vert }G_{b\beta
}^{+}\left( 0,z^{\prime };\varepsilon \right) \right] .
\end{equation}
Taking this equation at $z=0$ we find $G_{b\beta }^{+}\left( 0,z^{\prime
};\varepsilon \right) $ and finally $G_{b\beta }^{+}\left( z,z^{\prime
};\varepsilon \right) $ is given by

\begin{equation}
G_{b\beta \beta ^{\prime }}^{+}\left( z,z^{\prime };\varepsilon \right) =%
\frac{m^{\ast }}{i\hbar ^{2}k_{\beta }}\left[ e^{ik_{\beta }\left| z^{\prime
}-z\right| }+r_{\beta }e^{ik_{\beta }\left( \left| z^{\prime }\right|
+\left| z\right| \right) }\right] ,  \label{G+bar}
\end{equation}
where 
\begin{equation}
r_{\beta }=-\frac{im^{\ast }U}{\hbar ^{2}k_{\beta }+im^{\ast }U}=\cos
\varphi _{\beta }e^{i\varphi _{\beta }};  \label{r_beta}
\end{equation}
is the amplitude of reflected wave; 
\begin{equation}
\varphi _{\beta }\left( \varepsilon \right) =\arcsin \left[ \frac{1}{\sqrt{%
1+\left( m^{\ast }U/\hbar ^{2}k_{\beta }\right) ^{2}}}\right] .  \label{phi}
\end{equation}
The amplitude $t_{\beta }$ of transmitted wave can be evaluated through the $%
r_{\beta }$ from the condition of continuity of electron wave function at $%
z=0$%
\begin{equation}
t_{\beta }=r_{\beta }+1=\frac{\hbar ^{2}k_{\beta }}{\hbar ^{2}k_{\beta
}+im^{\ast }U}=i\sin \varphi _{\beta }e^{i\varphi _{\beta }}.  \label{tb}
\end{equation}
The same functions $r_{\beta }$ and $t_{\beta }$ can be found from the
solution of the one dimensional Schr\"{o}dinger equation of a system with $%
\delta -$function barrier $U\delta \left( z\right) $ \cite{Flugge}.

The Eq.(\ref{G_eq}) can be solved exactly for any finite number of defects.
For that the Eq.(\ref{G_eq}) should be written in all points $\mathbf{r}_{i}$
of the defect positions and the functions $G\left( \mathbf{r}_{i}\mathbf{,r}%
^{\prime };\varepsilon \right) $ are found from the system of $i$ algebraic
equations.

By using the matrix elements (\ref{t_beta}) the conductance $G=\frac{dI}{dV} 
$ of the microconstriction as well as the shot noise $S\left( 0\right) $ can
be calculated.

\section{Voltage dependence of conductance and shot noise.}

To illustrate the effect of quantum interference of scattered electron waves
to the conductance and the shot noise we present the results for two cases:
(i) single impurity in the constriction with the barrier; (ii) two
impurities in the constriction without the barrier. For the first case the
Green's function takes the form: 
\begin{equation}
G\left( \mathbf{r,r}^{\prime };\varepsilon \right) =G_{b}\left( \mathbf{r,r}%
^{\prime };\varepsilon \right) +\frac{gG_{b}\left( \mathbf{r,r}%
_{1};\varepsilon \right) G_{b}\left( \mathbf{r}_{1}\mathbf{,r}^{\prime
};\varepsilon \right) }{1-gG_{b}\left( \mathbf{r}_{1}\mathbf{,r}%
_{1};\varepsilon \right) };  \label{G1}
\end{equation}
where $\mathbf{r}_{1}$ is the position of the impurity, a Green's function $%
G_{b}\left( \mathbf{r,r}^{\prime };\varepsilon \right) $ is defined by Eqs.(%
\ref{G+}), (\ref{G+bar}). If only two impurities are situated inside the
ballistic microconstriction, the solution of the Eq.(\ref{G_eq}) is 
\begin{gather}
G\left( \mathbf{r,r}^{\prime };\varepsilon \right) =G_{0}\left( \mathbf{r,r}%
^{\prime };\varepsilon \right) +\frac{1}{1-G_{1}\left( \mathbf{r}%
_{1};\varepsilon \right) G_{1}\left( \mathbf{r}_{2};\varepsilon \right)
G_{0}^{2}\left( \mathbf{r}_{1}\mathbf{,r}_{2};\varepsilon \right) }\times
\label{G12} \\
\sum_{i,k=1,2;i\neq k}\left\{ G_{1}\left( \mathbf{r}_{i};\varepsilon \right)
G_{0}\left( \mathbf{r,r}_{i};\varepsilon \right) \right. \left[ G_{0}\left( 
\mathbf{r}_{i}\mathbf{,r}^{\prime };\varepsilon \right) +\right.  \notag \\
\left. \left. G_{1}\left( \mathbf{r}_{k};\varepsilon \right) G_{0}\left( 
\mathbf{r}_{i}\mathbf{,r}_{k};\varepsilon \right) G_{0}\left( \mathbf{r}_{k}%
\mathbf{,r}^{\prime };\varepsilon \right) \right] \right\} ;  \notag
\end{gather}
where 
\begin{equation}
G_{1}\left( \mathbf{r}_{i};\varepsilon \right) =\frac{g}{1-gG_{0}\left( 
\mathbf{r}_{i}\mathbf{,r}_{i};\varepsilon \right) };
\end{equation}
and $G_{0}\left( \mathbf{r,r}^{\prime };\varepsilon \right) $ is the Green's
function of the ballistic microconstriction (\ref{G0}). Using the Eqs. (\ref
{G1}) and (\ref{G12}) it is easy to find the transmission probabilities $%
t_{\beta \beta ^{\prime }}$ (\ref{t_beta}).

At zero temperature the nonlinear conductance $G\left( V\right) $ and the
noise power $S\left( 0,eV\right) $ are given by the following expressions: 
\begin{equation}
G\left( V\right) =\sum_{\beta \beta ^{\prime }}\left[ \left| t_{\beta \beta
^{\prime }}\left( \varepsilon _{F}+\frac{eV}{2}\right) \right| ^{2}+\left|
t_{\beta \beta ^{\prime }}\left( \varepsilon _{F}-\frac{eV}{2}\right)
\right| ^{2}\right] .
\end{equation}
\begin{eqnarray}
S\left( 0,eV\right) &=&\sum_{\beta \beta ^{\prime }\beta ^{\prime \prime
}\beta ^{\prime \prime \prime }}\int\limits_{\varepsilon _{F}-\frac{eV}{2}%
}^{\varepsilon _{F}+\frac{eV}{2}}d\varepsilon \left\{ t_{\beta \beta
^{\prime }}^{\star }\left( \varepsilon \right) t_{\beta ^{\prime }\beta
^{\prime \prime }}\left( \varepsilon \right) \right. \left[ \delta _{\beta
^{\prime \prime }\beta ^{\prime \prime \prime }}\delta _{\beta ^{\prime
\prime \prime }\beta }-\right. \\
&&\left. \left. t_{\beta ^{\prime \prime }\beta ^{\prime \prime \prime
}}^{\ast }\left( \varepsilon \right) t_{\beta ^{\prime \prime \prime }\beta
}\left( \varepsilon \right) \right] \right\} .  \notag
\end{eqnarray}

To explain the analytical results we present the expansion of the
transmission coefficient (\ref{T}) on the constant of electron-impurity
interaction $g$ up to linear in $g$ term for the constriction with one
impurity in the point $\mathbf{r}_{1}=\left( \mathbf{R}_{1},z_{1}\right) $
and the barrier 
\begin{eqnarray}
T\left( \varepsilon \right) &=&\sum_{\beta }\left\vert t_{\beta }\right\vert
^{2}\left\{ 1-\right.  \label{T_g} \\
&&\left. \frac{2m^{\ast }g}{\hbar ^{2}k_{\beta }}\left\vert r_{\beta
}\right\vert \left\vert \psi _{\bot \beta }\left( \mathbf{R}_{1}\right)
\right\vert ^{2}\cos \left( 2k_{\beta }z_{1}+\varphi _{\beta }\right)
\right\} ;\varepsilon >\varepsilon _{\beta },  \notag
\end{eqnarray}
where $t_{\beta },$ $r_{\beta },$ and phase $\varphi _{\beta }$ are defined
by Eqs.(\ref{tb}), (\ref{r_beta}), and (\ref{phi}). This formula is correct
for $\frac{2m^{\ast }g}{\hbar ^{2}k_{\beta }}\ll 1,$ i.e. far from the end
of the step of conductance, where $k_{\beta }\rightarrow 0.$ The oscillatory
term in Eq.(\ref{T_g}) originates from the interference between directly
transmitted wave (trajectory 1 in Fig.1) and the wave, which is once
reflected by the barrier and after one reflection from the impurity passes
through the contact (trajectory 2 in Fig.1). The amplitude of the
oscillations depends on the local density of electron states $\nu _{\beta
}\left( \mathbf{R}_{1},\varepsilon \right) =m^{\ast }\left\vert \psi _{\bot
\beta }\left( \mathbf{R}_{1}\right) \right\vert ^{2}/\left( \hbar
^{2}k_{\beta }\left( \varepsilon \right) \right) $ in the point, in which
the impurity is situated. In certain points $\nu _{\beta }\left( \mathbf{R}%
,\varepsilon \right) $ can be equal to zero and the defect located near such
point contribute a little to the oscillatory addition of $\beta $th mode to
the $T\left( \varepsilon \right) .$ In particular, impurities on the surface 
$\mathbf{R}=\mathbf{R}_{s}$ do not influence to the oscillations of $T\left(
\varepsilon \right) ,$ because $\psi _{\bot \beta }\left( \mathbf{R}%
_{s}\right) =0.$ As a result of the reflection from the barrier the
oscillations have the additional phase $\varphi _{\beta }$. Its dependence
on the energy $\varepsilon $ leads to nonperiodicity of oscillations of
function $T\left( \varepsilon \right) .$ The Eq. (\ref{T_g}) enables to
follow the changing in the amplitude of oscillation with changing of the
contact diameter. If the diameter is increased and goes to the end of the
conductance step, the energy of the transverse quantum mode $\varepsilon
_{\beta }$ is decreased (see, for example, the Eq. (\ref{e_mn}) for
cylindrical geometry). The wave number $k_{\beta }$ (\ref{k_beta}) is
increased and according the Eq. (\ref{r_beta}) the modulus of the reflection
probability $\left\vert r_{\beta }\right\vert $ is decreased. In opposite
situation (the radius is decreased) the decreasing of $k_{\beta }$ is a
reason for the decreasing of the transmission probability $\left\vert
t_{\beta }\right\vert $ (\ref{tb}). In both cases the amplitude of
oscillations of $T\left( \varepsilon \right) $ is decreased.

The similar expansion of $T\left( \varepsilon \right) $ for the constriction
with two defects in points $\mathbf{r}_{1}=\left( \mathbf{R}%
_{1},z_{1}\right) $ and $\mathbf{r}_{2}=\left( \mathbf{R}_{2},z_{2}\right) $
without barrier is 
\begin{eqnarray}
T\left( \varepsilon \right) &=&\sum_{\beta \beta ^{\prime }}\left\{ \delta
_{\beta \beta ^{\prime }}-2\left( \frac{m^{\ast }g}{\hbar ^{2}}\right) ^{2}%
\frac{1}{k_{\beta }k_{\beta ^{\prime }}}\sum_{i=1,2}\left[ \left\vert
A_{\beta \beta ^{\prime }}^{(ii)}\right\vert ^{2}+\right. \right. \\
&&\left. \left. \func{Re}\sum_{i\neq j=1,2}A_{\beta \beta ^{\prime
}}^{(ii)}A_{\beta \beta ^{\prime }}^{(jj)}\exp \left( \left( k_{\beta
}+k_{\beta ^{\prime }}\right) \left( z_{j}-z_{i}\right) +\varphi _{\beta
}+\varphi _{\beta ^{\prime }}\right) \right] \right\} ,  \notag \\
\varepsilon &>&\varepsilon _{\beta },\varepsilon _{\beta ^{\prime }};  \notag
\end{eqnarray}
where 
\begin{equation}
A_{\beta \beta ^{\prime }}^{(ii)}=\psi _{\bot \beta }\left( \mathbf{R}%
_{i}\right) \psi _{\bot \beta ^{\prime }}^{\ast }\left( \mathbf{R}%
_{i}\right) .
\end{equation}
The last term in square brackets describes the interference effect between
trajectory 3 in Fig.1 and trajectory 4, which corresponds to two scattering
by different impurities, and nonmonotonically depends on the energy $%
\varepsilon .$ The discussed above the energy dependence of the transmission
coefficient $T\left( \varepsilon \right) $ manifests itself in nonmonotonic
dependence of conductance and shot noise on the applied bias $eV.$

The general expression for the components $t_{\beta \beta ^{\prime }}\left(
\varepsilon \right) $ (\ref{t_beta}) calculated by using the Green's
functions (\ref{G1}) and (\ref{G12}) takes into account a multiple electron
scattering by the impurities and barrier. It is valid for any values of
parameters. Below we illustrate such situation presenting the plots for the
voltage dependencies of conductance and shot noise for some values of
parameters, which may be related to experiments.

\bigskip For numerical calculations we used the model of cylindrical
channel, for which in formulas (\ref{psi}) and (\ref{energ}) 
\begin{eqnarray}
\psi _{\bot \beta }\left( \rho ,\varphi \right) &=&\frac{1}{\sqrt{\pi }%
RJ_{m+1}\left( \gamma _{mn}\right) }J_{m}\left( \gamma _{mn}\frac{\rho }{R}%
\right) e^{im\varphi };  \label{e_mn} \\
\varepsilon _{mn} &=&\frac{\hbar ^{2}\gamma _{mn}^{2}}{2m^{\ast }R^{2}}.
\end{eqnarray}
Here we used the cylindrical coordinates $\mathbf{r=}\left( \rho ,\varphi
,z\right) ;$ $\gamma _{mn}$ is $n-th$ zero of Bessel function $J_{m}.$ Also
we introduce dimensionless parameters 
\begin{equation}
\widetilde{g}=\frac{m^{\ast }g}{\pi R^{2}\hslash ^{2}k_{F}};\quad \widetilde{%
U}=\frac{m^{\ast }U}{\hslash ^{2}k_{F}},  \label{dim_less}
\end{equation}
where $k_{F}$ is the Fermi wave vector. We have performed the calculations
for $\widetilde{g}=1,\ \widetilde{U}=0.5$
For such values of these parameters the amplitude of conductance
oscillations is of the order the amplitude, which was observed in Ref.\cite
{Kempen}. For the value of the radius $2\pi R=2.9\lambda _{F}$ (one mode
channel) the first energy level $\varepsilon _{0,1}<\varepsilon _{F}$ is
comparatively far from the Fermi energy and for $2\pi R=3.45\lambda _{F}$
this level is closed to $\varepsilon _{F}.$ For the larger value of radius $%
\left( 2\pi R=5\lambda _{F}\right) $ there are two opened quantum modes with
energies $\varepsilon _{0,1}$, $\varepsilon _{\pm 1,1}<\varepsilon _{F}.$ To
illustrate different reasons \ of appearance of conductance oscillations, in
Fig.2 and Fig.3 we show the dependences of the conductance on the applied
voltage for the channel without the barrier $\left( U=0\right) $ containing
two impurities and for the channel with the barrier and single impurity.
From comparison of the different curves in Figs. 2,3 we observe that the
amplitude of conductance oscillations is decreased for radius value ($2\pi
R=3.45\lambda _{F}$) corresponding the end of first step of conductance. In
Fig.4 and Fig.5 the voltage dependences of noise power are plotted. We
remark that, as seen from Fig.4, for the one mode channel the shot noise is
the strongly nonmonotonic function of $V.$ As well as for the conductance,
the amplitude of the oscillations of the shot noise is decreased near the
end of the first step ($2\pi R=3.45\lambda _{F}$). For the two mode channel
the $S(V)$ is almost linear function that can be explained by the effect of
a superposition of oscillations with a different periods. In the contact
with the barrier the main part of the shot noise $S_{0}\left( V\right) $
originates from the electron reflection from the barrier potential ($S\left(
V\right) =S_{0}\left( V\right) ,$ if $g=0),$ and it is the monotonic
function on $V.$ The small nonlinearity of this function arises from the
energy dependence transmission probability. The interference of electron
wave in the presence of defect leads to nonmonotonic additions, which we
show in Fig.5.

\section{Conclusion}

We have studied theoretically the voltage dependence of the conductance $G$
and the shot noise power $S$ in the quantum microconstriction in the form of
long channel (quantum wire). The effect of quantum interference of electron
waves, which are scattered by single defects and the potential barrier
inside the constriction, is taken into account. In the framework of the
model we have obtained the analytical solution of the problem and found the
dependencies of the $G$ and $S$ on such important parameters as the
constriction diameter, the constant of electron-impurity interaction, the
amplitude of the barrier potential and positions of impurities. In the
general case these dependencies are complex and defined by the expression of
transmission probability $t_{\beta \beta ^{\prime }}$ (\ref{t_beta}) by
means of Green's functions (\ref{G1}), (\ref{G12}). For the small constant $%
g $ of electron-impurity interaction and far from the step of conductance
the part of the total transmission coefficient $T(\varepsilon )$(\ref{T_g}),
which is due to the interference effect, is proportional to $g$ and to the
amplitude of the reflected from the barrier wave $r_{\beta }$ (see, Eq.(\ref
{r_beta})). As a result of that, at small $g$ and $U$ the interference part
of conductance and shot noise is proportional to $gU$ or $g^{2}$ (for $U=0$)
for any number of defects.

We have shown that the conductance and noise are oscillatory functions on
the applied bias $V$ and come to the conclusion that the experimentally
observed suppression of conductance oscillations \cite{Ludoph} could be due
to the energy dependence of the transmission probability of electrons
through the constriction. In the framework of our model this suppression of
conductance oscillations can be explained in the following way: The
oscillatory part of conductance is decreased with the decreasing of
amplitude $r_{\beta }$ of reflected from the barrier wave. The reflection
probability $r_{\beta }$ from the barrier has the minimal value, if the
energy of quantum mode $\varepsilon _{\beta }$ is close to $\varepsilon
_{\beta }\lesssim \varepsilon _{F}$ . It is demonstrated that in the one
mode constriction containing only impurities the shot noise power is a
strongly nonlinear function on $V.$ \ In the contact with the barrier the
almost linear dependence $S\left( V\right) $ has small oscillatory addition.

We acknowledge fruitful discussion with A.N. Omelyanchouk.

\newpage

\begin{figure*}[ht!]
\centerline{\includegraphics[width=14.0cm]{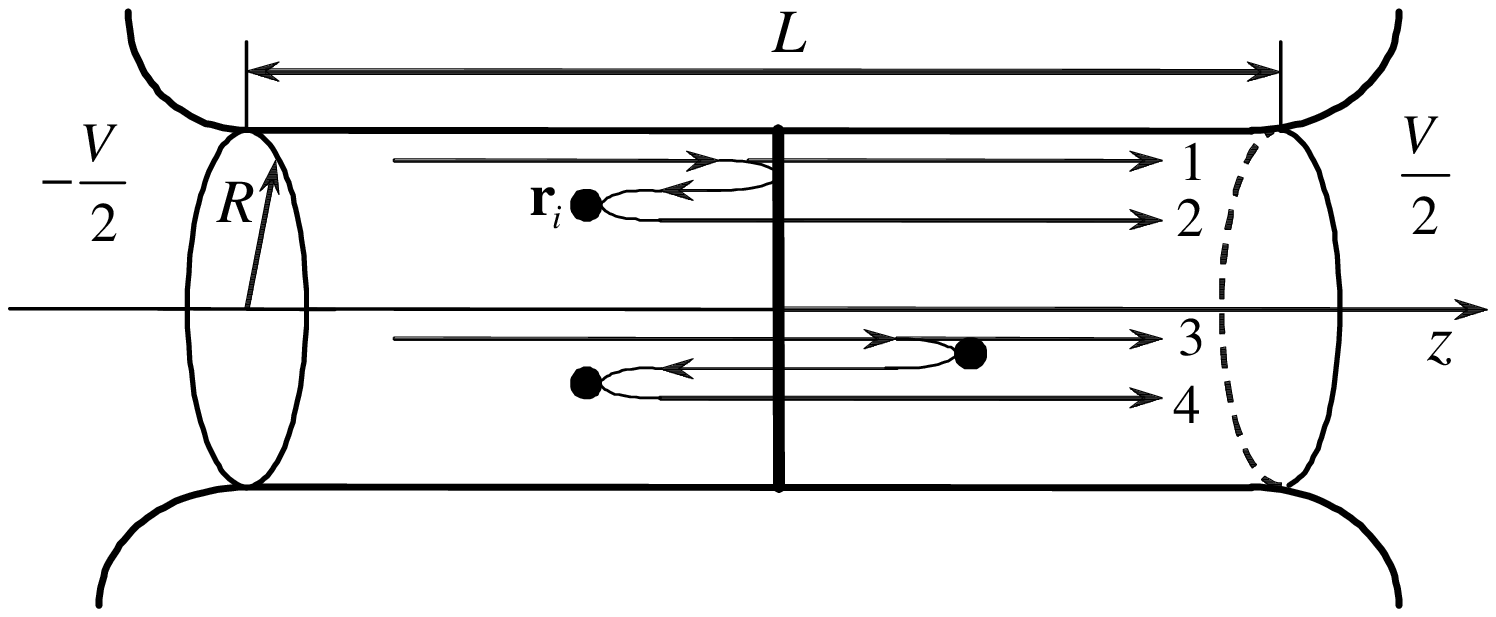}}
\caption{ The model of quantum constriction in the form of long channel
adiabatically connecting with bulk metallic reservoirs. The trajectories
(1-4) of electrons , which are scattered by the defects and a barrier are
shown schematically.}
\end{figure*}

\begin{figure*}[ht!]
\centerline{\includegraphics[width=14.0cm]{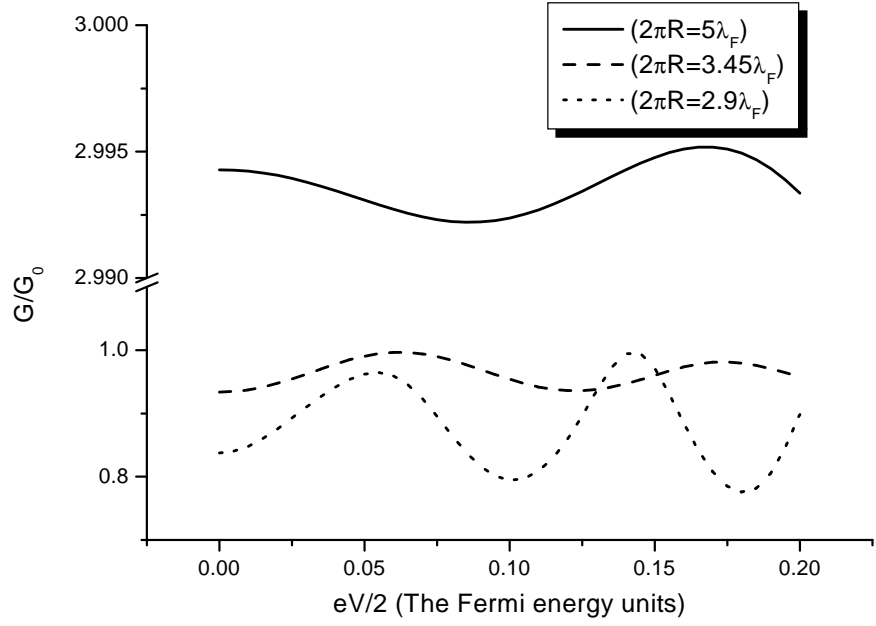}}
\caption{The dependences of the conductance on the applied voltage for the
channel containing two impurities for different values of radius; impurity
positions are $2\protect\pi \protect\rho _{1}=0.3\protect\lambda _{F}$ and $2%
\protect\pi \protect\rho _{2}=0.4\protect\lambda _{F},$ $2\protect\pi \left(
z_{1}-z_{2}\right) =35\protect\lambda _{F}.$ }
\end{figure*}

\begin{figure*}[ht!]
\centerline{\includegraphics[width=14.0cm]{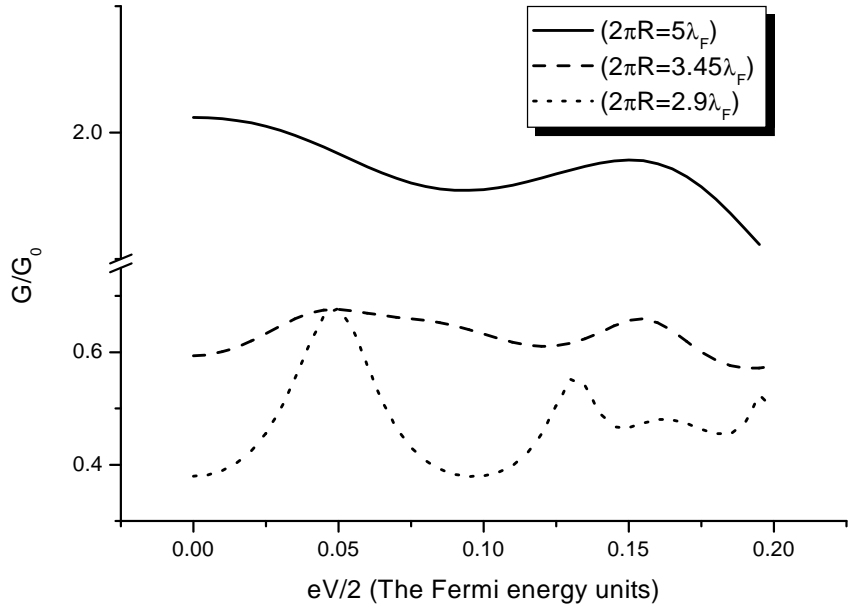}}
\caption{The dependences of the conductance on the applied voltage for the
channel containing the single impurity and the barrier for different values
of radius; the impurity position is $2\protect\pi \protect\rho _{1}=0.3%
\protect\lambda _{F},$ $2\protect\pi z_{1}=35\protect\lambda _{F}.$ }
\end{figure*}

\begin{figure*}[ht!]
\centerline{\includegraphics[width=14.0cm]{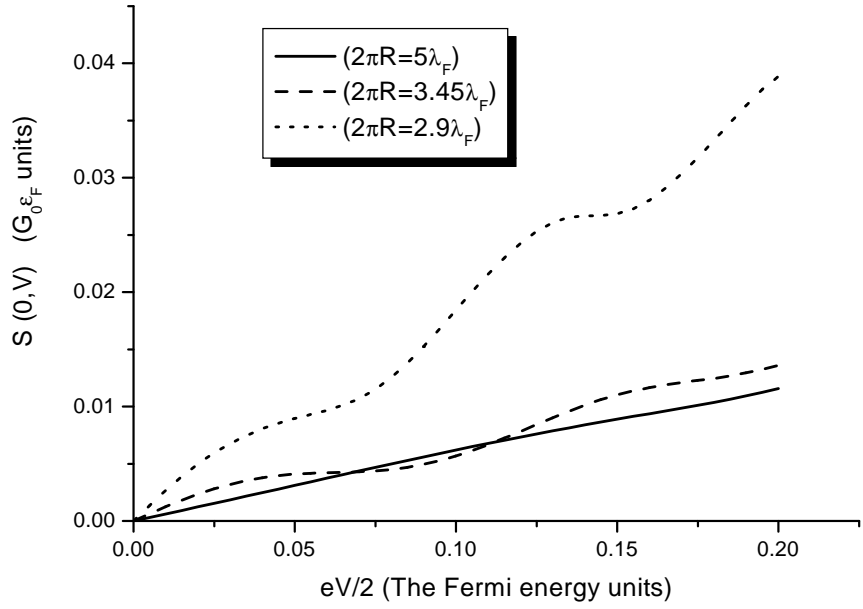}}
\caption{The voltage dependences of noise power on the applied voltage for
the channel containing two impurities for different values of radius;
impurity positions are $2\protect\pi \protect\rho _{1}=0.3\protect\lambda
_{F}$ and $2\protect\pi \protect\rho _{2}=0.4\protect\lambda _{F}$, $2%
\protect\pi \left( z_{1}-z_{2}\right) =35\protect\lambda _{F}.$ }
\end{figure*}

\begin{figure*}[ht!]
\centerline{\includegraphics[width=14.0cm]{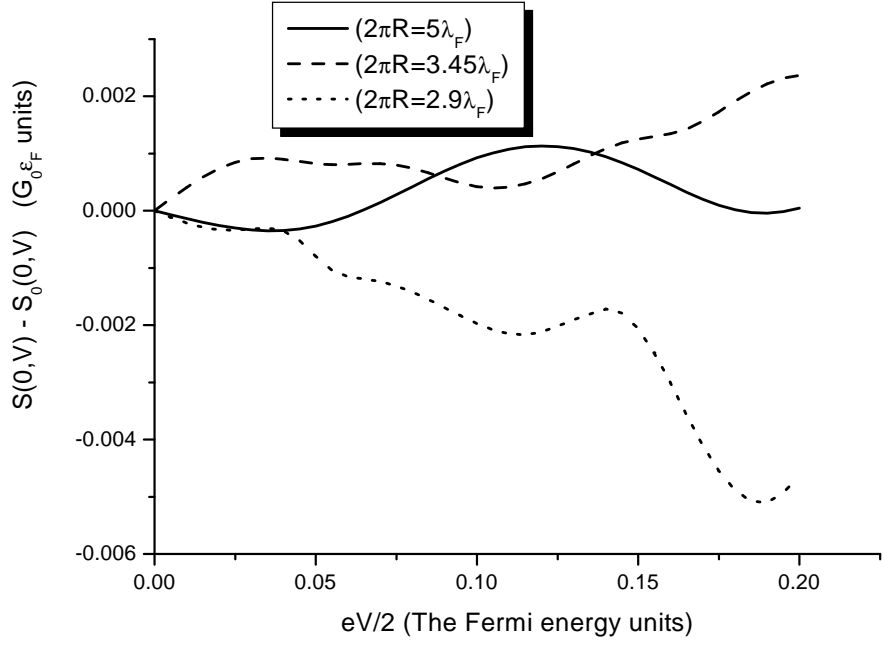}}
\caption{The voltage dependences of the nonmonotonic part of noise power on
the applied voltage for the channel containing the single impurity and the
barrier for different values of radius; the impurity position is $2\protect%
\pi \protect\rho _{1}=0.3\protect\lambda _{F},$ $2\protect\pi z_{1}=35%
\protect\lambda_{F}.$}
\end{figure*}

\end{document}